\documentclass{llncs}
\usepackage{etex}
\usepackage{amssymb}
\usepackage{graphicx}
\usepackage[ruled,vlined,linesnumbered]{algorithm2e}
\usepackage{caption,wrapfig}
\usepackage{mathrsfs}
\usepackage{url}
\usepackage{amsfonts}
\usepackage{times}
\usepackage{pgf}
\usepackage{tikz}
\usetikzlibrary{arrows,automata,positioning}
\usepackage[latin1]{inputenc}
\usepackage{verbatim}
\usepackage{makecell}
\usepackage{booktabs}
\usepackage{adjustbox}
\usepackage{listings}
\usepackage{caption}
\usepackage{subcaption}
\usepackage{moresize}

\usepackage{array}

\newcolumntype{H}{>{\setbox0=\hbox\bgroup}c<{\egroup}@{}}

\newcommand{\myit}[1]{\textit{#1}}
\newcommand{\btt}[1]{{\ensuremath{\tt #1}}}

\newcommand{\form}[1]{\ensuremath{#1}}
\newcommand{\code}[1]{{\small {\ensuremath{\tt #1}}}}

\newcommand{\heap}{\ensuremath{\kappa}}
\newcommand{\pure}{\ensuremath{\pi}}

\def\D{\Delta}
\newcommand{\constr}{\ensuremath{\Phi}}
\newcommand{\sepnode}[3]{\ensuremath{#1{\mapsto}#2(#3)}}
\newcommand{\sepnodeF}[3]{\ensuremath{{#1}{\pto}#2(#3)}}
\newcommand{\seppredF}[2]{\ensuremath{#1(#2)}}
\newcommand{\seppred}[2]{\ensuremath{#1(#2)}}
\newcommand{\pto}{{\scriptsize\ensuremath{\mapsto}}}
\newcommand{\setvars}[1]{\ensuremath{\bar{#1}}}

\newcommand{\emp}{\btt{emp}}
\def\sep{\code{*}}
\newcommand{\nil}{{\code{null}}}
\def\a{a}
\newcommand{\anon}{\ensuremath{\_\,}}
\def\true{\code{true}\,}
\def\false{\code{false}\,}

\newcommand{\stool}{SLearner}

\author{Long H. Pham\inst{1}\thanks{Corresponding author. Email: longph1989@gmail.com} \and Jun Sun\inst{2} \and Quang Loc Le\inst{3}}
\institute{Singapore University of Technology and Design, Singapore 
\and {Singapore Management University, Singapore}
\and School of Computing \& Digital Technologies, Teesside University, UK
}

\begin{document}
\title{Compositional Verification of Heap-Manipulating Programs through
Property-Guided Learning}


\maketitle

\begin{abstract}
Analyzing and verifying heap-manipulating programs automatically is challenging. A key for fighting the complexity is to develop compositional
methods. 
For instance, many existing verifiers for heap-manipulating programs require user-provided specification for each function in the program in order to decompose the verification problem. The requirement, however, often hinders the users from applying such tools. To overcome the issue, we propose to automatically learn heap-related program invariants in a property-guided way for each function call. The invariants are learned based on the memory graphs observed during test execution and
improved through memory graph mutation. We implemented a prototype of our approach and integrated it with two existing program verifiers. The experimental results show that our approach enhances existing verifiers effectively in automatically verifying complex heap-manipulating programs with multiple function calls. 
\end{abstract}

\section{Introduction}
Analyzing and verifying heap-manipulating programs (hereafter heap programs) automatically 
is 
challenging~\cite{Reynolds:LICS02}. Given the complexity, the key is to develop compositional
methods which 
allow us to
decompose a complex 
problem into smaller manageable ones. One successful example is the Infer static analyzer~\cite{infer}, which applies techniques like bi-abduction for local reasoning~\cite{DBLP:conf/csl/OHearnRY01}
to infer a specification for each function in a program
to be analyzed.

While Infer generates function specifications for identifying certain classes of program errors, we aim to develop compositional methods for the more challenging task of verifying heap programs with data structures.
%
In recent years, there have been multiple tools developed to verify heap programs in a compositional way, including Dafny~\cite{Leino:LPAR:2010}, GRASShopper~\cite{Piskac:CAV:2014,Piskac:TACAS:2014} and HIP~\cite{Chin:SCP:2012}. These tools are, however, far from being applicable to real-world complex programs. One reason is that substantial user effort is needed. In particular, besides providing a specification to verify against, users must provide auxiliary specification to decompose the verification problem. For instance, Dafny, GRASShopper and HIP all require users to provide a specification for each function used in the program. Writing the function specification is highly non-trivial.
It is thus desirable to develop approaches for verifying heap programs in a compositional way which requires minimum user effort.

In this work, we propose to automatically generate function specifications for compositional verification of heap programs. Our approach differs from existing approaches like Infer in three ways. Firstly, because our goal is to verify the correctness of heap programs with \emph{data structures},
our approach generates more expressive function specifications than those generated by Infer. 

Secondly, we learn a specification of each function call (rather than each function) in a property-guided way. For instance, assume that we have the following verification problem 
(expressed in the form of a Hoare triple)
$\{pre\} func(); func(); \{post\}$
where $pre$ is a precondition, $post$ is a postcondition and $func(); func()$ are two consecutive calls of the same function. We automatically generate a program invariant $inv$ after the first function call and before the second function call. As a result, we generate the specification $\{pre\}func()\{inv\}$ for the first function call and the specification $\{inv\}func()\{post\}$ for the second function call. The (smaller) problems of verifying these two Hoare triples thus replace the problem of verifying the original Hoare triple.

Thirdly, our invariant generation method is based on a novel technique, namely, a combination of classification and memory-graph mutation.
We start with generating multiple random test cases (based on existing methods~\cite{randoop}). We then instrument the program and execute the test cases to obtain values of multiple features which are related to the memory graphs before and after each function call in the program. The obtained feature vectors are labeled according to the testing results (i.e.,
whether the postcondition is satisfied or not). Then we apply a classification algorithm~\cite{DBLP:journals/cacm/Valiant84} to find an invariant that separates the feature vectors with different labels. The invariant is an arbitrary boolean formula of the features, which is then used to decompose the verification problem.

There are two technical challenges which we must solve in order to make the above approach work. First, what features of the memory graphs shall we use? 
In this work, we adopt an expressive specification language for heap programs which combines separation logic, user-defined inductive predicates and arithmetic~\cite{Chin:SCP:2012,Ishtiaq:POPL01,DBLP:conf/cav/LeSC16,Reynolds:LICS02}. We then define a set of features based on the specification language.
In addition, our approach allows users to define their own features.
Secondly, how do we solve the problem of the lack of labeled samples, i.e., the test cases which we learn from may be limited. To overcome the problem, 
we mutate the memory graphs according to the learned invariant to validate whether the learned invariant is correct.
We refine the invariant based on the validation result (if necessary) and repeat the process until the invariant is validated.

We implement our idea in a prototype, called \stool, which takes a program to be verified as input, generates multiple invariants and outputs a set of decomposed verification tasks. 
We integrate \stool~with two existing state-of-the-art verifiers for heap programs, i.e., GRASShopper and HIP. Experiments are then performed on 110 programs manipulating 10 challenging data structures. The experimental results show that, 
enhanced with our approach, both GRASShopper and HIP are
able to successfully verify programs with multiple function calls without user-provided function specifications.

The novelty of our work is in learning heap-related specification in a property-guided way
 and applying graph mutation to improve the learning process. The rest of the paper is organized as follows. Section~\ref{anexample} presents an illustrative example.
Section~\ref{preliminary} presents the details of our approach.
Section~\ref{evaluation} evaluates our approach.
Section~\ref{related} reviews related work.
Finally, Section~\ref{conclusion} concludes.

\begin{figure}[t]
{\scriptsize\[
\begin{minipage}{0.48\textwidth}
\[
\begin{array}{ll}
1 & \code{public~void~main(int~m,~int~n)~\{}  \\
2 & \code{\quad //precondition: m \leq n}  \\
3 & \code{\quad Node~x~=~createSLL(m);}  \\
4 & \code{\quad Node~y~=~createSLL(n);} \\
5 & \code{\quad getSum(x, y);} \\
6 & \code{\quad //postcondition: \form{\code{sll}(x,\anon)\sep \code{sll}(y,\anon)}} \\
7 & \code{\}} \\
8 & \code{   private~Node~createSLL(int~n)~\{}  \\
9 & \code{ \quad       if~(n <= 0)~return~null;}  \\
10 & \code{  \quad     else \{}  \\
11 & \code{  \quad  \quad       Node~x~=~new~Node(n,~null);}  \\
12 & \code{  \quad  \quad       x.next~=~createSLL(n-1);}  \\
13 & \code{  \quad  \quad       return~x;}  \\
\end{array}
\]
\end{minipage}
\begin{minipage}{0.48\textwidth}
\[
\begin{array}{ll}
14 & \code{  \quad     \}}  \\
15 & \code{   \}}  \\
16 & \code{   private~int~getSum(Node~x,~Node~y) \{} \\
17 & \code{\quad       int~sum~=~0;} \\
18 & \code{\quad       if~(x~!=~null)~\{} \\
19 & \code{\quad\quad           sum~+=~x.data + y.data;} \\
20 & \code{\quad\quad           sum~+=~getSum(x.next,~y.next);} \\
21 & \code{\quad       \}} \\
22 & \code{\quad       return~sum;} \\
23 & \code{   \}}
\end{array}
\]
\end{minipage}
\]}
\caption{An illustrative example}
\label{anexample0}
\end{figure}

\section{An Illustrative Example}
In this section, we illustrate 
 our approach 
 with an example. The program is shown as function \form{\code{main}} in Fig.~\ref{anexample0}, where function \form{\code{createSLL}(n)} returns a singly-linked list with length \form{n} and function \form{\code{getSum}(x,y)} returns the sum of the \form{\code{data}} in two disjoint singly-linked list objects (pointed to by the two pointers \form{x} and \form{y}). Note that both functions are recursively defined. The precondition and postcondition are shown at line 2 and 6 respectively.
They are specified in an assertion language based on separation logic
(refer to details in Section~\ref{preliminary}). The precondition is self-explanatory. The postcondition \form{\code{sll}(x,\anon)\sep \code{sll}(y,\anon)} intuitively means that \form{x} and \form{y} are two disjoint singly-linked list objects, i.e., \form{\code{sll}(x,n)} is an inductive predicate denoting that \form{x} is a singly-linked list object with \form{n} nodes, and \form{\sep} is the separating conjunction predicate specifying the disjointness in separation logic.
Besides the postcondition, we assume that memory safety is always implicitly asserted and thus must be verified. For instance, we aim to verify that \form{\code{x.data}} at line 19 would not result in null-pointer de-referencing.

Our experiment shows that state-of-the-art 
verifiers like GRASShopper and HIP cannot verify this program. Only after specifications for both functions \form{\code{createSLL}} and \form{\code{getSum}} are provided manually, the program is verified. On one hand, providing a specification for every function called by the given program is highly nontrivial. On the other hand, part of the function specification may be irrelevant to verifying the given program. For an extreme example, if we change the postcondition of the program shown in Fig.~\ref{anexample0} to \form{\code{true}}, a complete specification for singly-linked lists would not be necessary to verify the program.

\begin{table*}[t]
\centering
\scriptsize
\caption{Collected feature vectors and labels}
\begin{tabular}{ c | c |  c | c | c | c}
 & \form{\code{is\_sll}(x)} &
  \form{\code{is\_sll}(y)} & 
  \form{\code{is\_sll}(x) \wedge \code{is\_sll}(y) \wedge \code{sep}(x,y)}
  &
 \form{\code{len\_sll}(x)} $\leq$ \form{\code{len\_sll}(y)} & label \\
\hline
m=1, n=0 & true & true & true & false & negative \\
\hline
m=0, n=1 & true & true & true & true & positive \\
\hline
\end{tabular}
\label{summary}
\end{table*}

Our approach is to automatically learn a just-enough invariant before and after each function call so that we can verify the program in a compositional way. For this example, we learn two invariants: \form{\code{inv_1}} right after the first function call at line 3 and \form{\code{inv_2}} right after the second function call at line 4.
Next, we verify the program by verifying the following three Hoare triples: $\{m \leq n\}\code{createSLL(m)}\{\code{inv_1}[\code{res}/x]\}$; $\{\code{inv_1}\}\code{createSLL(n)}\{\code{inv_2}[\code{res}/y]\}$; and
 $\{\code{inv_2}\}\code{getSum}\{\code{sll}(x,\anon)\sep \code{sll}(y,\anon)\}$ with
 \form{\code{res}} is a special variable for the return value of a function and \form{\code{inv_1}[\code{res}/x]} is a substitution
of all variable \form{x} in \form{\code{inv_1}}
by variable \form{\code{res}}.
As the program in each Hoare triple involves only one function, existing
verifiers like GRASShopper and HIP can automatically verify the Hoare triples.


To learn \form{\code{inv_1}} and \form{\code{inv_2}}, we instrument the program to collect a set of features at the learning points and collect their values during test executions.
For instance, Table~\ref{summary} shows a few of the features and their values for the above program after line 4 for learning \form{\code{inv_2}}. The first row shows the features and the second and third rows show the values of the features given two test cases $\{m {=} 1, n {=} 0\}$ and $\{m {=} 0, n {=} 1\}$ respectively. The features are designed based on our assertion language.
In particular, feature \form{\code{len\_sll}(x)} is a numeric value denoting the length of a singly-linked list \form{x} which is extracted based on the user-defined predicate \form{\code{sll}}; feature \form{\code{is\_sll}(x)} denotes whether \form{x} points to a singly-linked list, and feature \form{\code{sep}(x,y)}
denotes whether \form{x} and \form{y} are disjoint in the heap. We label each feature vector
with either $negative$ or $positive$, where $negative$ means that a memory error is generated, the postcondition is violated, or the test case likely runs into infinite loop (i.e., it does not stop after certain time units); and $positive$ means otherwise. 


%
Next, we apply a classification algorithm~\cite{DBLP:journals/cacm/Valiant84} to generate a predicate which separates the $positive$ and $negative$ feature vectors. The predicate takes the form of an arbitrary boolean formula of the features. Given the feature vectors in Table~\ref{summary}, the generated predicate is: \form{\code{len\_sll}(x)} $\leq$ \form{\code{len\_sll}(y)}. Although this predicate is an invariant after line 4, it is not strong enough to verify the postcondition. This is in general a problem due to having a limited number of test cases. To solve the problem, we systematically mutate the memory graphs obtained during the test executions to obtain more labeled feature vectors with the aim to improve the predicate (see details in Section~\ref{active}). In our example, with the additional feature vectors, the classification algorithm generates the following predicate for \form{\code{inv_2}}.
\[
\footnotesize
\begin{array}{l}
(\form{\code{is\_sll}(x)} \wedge \form{\code{is\_sll}(y)}~{\wedge}~\form{\code{sep}(x,y)}~{\wedge}~x {=} \code{null})~{\vee} \\
 \quad (\form{\code{is\_sll}(x)} \wedge \form{\code{is\_sll}(y)}~{\wedge}~\form{\code{sep}(x,y)} ~{\wedge}~ \form{\code{len\_sll}(x)} \leq \form{\code{len\_sll}(y)})
\end{array}
\]
\noindent We obtain $x{=}\code{null} \vee (\form{\code{is\_sll}(x)}~{\wedge}~\form{\code{len\_sll}(x)} \leq n)$ similarly for \form{\code{inv_1}} after line 3 .


Afterwards, \form{\code{inv_1}} and \form{\code{inv_2}} are translated into the formulas in our assertion language. 
Note that the translation is straightforward since the features are designed based on the assertion language.
The last step is to verify three verification problems.
This is done using state-of-the-art verifiers for heap programs. For instance, HIP solves the three verification problems automatically, which verifies the program.

{For efficiency,
in the verification step we perform the following two simplifications.
First, for dead code detection, we invoke a separation logic
 solver (e.g., the one presented in \cite{DBLP:conf/cav/LeSC16,Le:CAV:2017}) 
to check the satisfiability of inferred invariant.
Secondly,
we identify and eliminate the frame
of a Hoare triple before sending them to the verifiers.
For example, for the Hoare triple
$\{\code{inv_1}\}\code{createSLL(n)}\{\code{inv_2}[\code{res}/y]\}$,
we find that \form{x} has not been accessed
by the code,
  the occurrences of the singly-linked list \form{x}
in both the precondition and postcondition of the triple
are eliminated before sending it to the verifiers.}

 \label{anexample}
\section{Our Approach} \label{preliminary} 

\subsection{Problem Definition}
Our input is a Hoare triple $\{pre\}prog\{post\}$, where $pre$ is a precondition, $post$ is postcondition and $prog$ is a heap program which may invoke other functions. One example is the function \form{\code{main}} shown in Fig.~\ref{anexample0}. The precondition and postcondition are in an expressive specification language previously developed in~\cite{Chin:SCP:2012,Ishtiaq:POPL01,DBLP:conf/cav/LeSC16,Reynolds:LICS02}. The language supports separation logic, inductive predicates and Presburger arithmetic~\cite{pred}, which is shown to be expressive to capture many properties of heap programs.
 \begin{figure}[t]
{\footnotesize\[
\begin{minipage}{0.48\textwidth}
\[
\begin{array}{lll}
\constr & {::=} & \D ~|~ \constr_1 ~{\vee}~ \constr_2 \\
\D & {::=} & \exists \bar{v}{\cdot}~(\heap\wedge\pure) \\
 \heap & {::=} & \emp ~|~ \sepnodeF{r}{c}{\setvars{t}} ~|~ \seppredF{\code{P}}{\setvars{t}}~|~\heap_1 \sep \heap_2 \\
 \pure & {::=} & \true ~|~ \false ~|~  
 \phi \mid 
 \pure_1 \wedge \pure_2 \\ 
\end{array}
\]
\end{minipage}
\begin{minipage}{0.48\textwidth}
\[
\begin{array}{lll}
\phi & {::=} & \true ~|~ \myit{i}  \mid 
 v{=}\nil  
  ~|~ \phi_1{\wedge}\phi_2 \\
\myit{i} & ::= & \a_1 {=} \a_2 \mid \a_1 {\leq} \a_2 \\
\a & ::= & \!k \mid v \mid k{\times}\a \mid \a_1 \!+\! \a_2 \mid - \a \\ 
\end{array}
\]
\end{minipage}
\]}
 \caption{Syntax: where $c$ is a data type; $k$ is an integer value; $t_i$, $v$, $r$ are variables; and $\setvars{t}$ is a sequence of variables} 
 \label{spec.fig} 
 \end{figure}

The syntax of the language is presented in Fig.~\ref{spec.fig}.
In general, a predicate \form{\constr} in this language is a disjunction of
multiple symbolic heaps.
A symbolic heap \form{\D} is an existentially quantified conjunction of a heap formula \form{\heap} (i.e., a predicate constraining the memory structure)
and a pure formula \form{\pure} (i.e., a predicate constraining numeric variables).
A heap formula \form{\heap} is an empty heap predicate \form{\emp}, a points-to predicate \form{\sepnodeF{r}{c}{\setvars{t}}}
(where \form{r} is its root variable), a user-defined predicate \form{\seppredF{\code{P}}{\setvars{t}}}, or a spatial conjunction of two heap formulas \form{\heap_1} \form{\sep} \form{\heap_2}. User-defined predicates are defined in the same language.
A pure formula  \form{\pure} can be \form{\true}, an (in)equality on variables, a
Presburger arithmetic formula, negation of a formula,
 or their conjunction.
 We
 refer the readers to~\cite{pred} for details on Presburger arithmetic.
We note that \form{v_1 {\neq} v_2} (resp. \form{v {\neq} \nil}) is used to denote \form{\neg (v_1{=}v_2)}
 (resp. \form{\neg(v{=}\nil)}) and we may use $\_$ to indicate ``don't care'' values.

For instance, the following predicate \form{\seppred{\code{sll}}{x{,}n}} defines a singly-linked list (with a root-pointer \form{x} and size \form{n}),
which is used in the illustrative example.
\[
\footnotesize
\begin{array}{ll}
\form{\seppred{\code{sll}}{x{,}n} \equiv &
(\emp~{\wedge}~x{=}\nil~{\wedge}~{n{=}0})}\\
 & {\vee}~
\form{({\exists} ~q{,}n_1 {\cdot}~\sepnode{x}{Node}{\anon{,}q}~{\sep}~
\seppred{\code{sll}}{q{,}n_1}~{\wedge}~n{=}n_1{+}1)}\\
\end{array}
\]

Our problem is to automatically verify the Hoare triple.
Different from existing approaches, we
aim to 
do that in a compositional way without user-provided function specifications.

\subsection{Test Generation and Code Instrumentation}
Given $\{pre\}prog\{post\}$, we first automatically generate a test suite $S$ using existing test case generation methods like~\cite{randoop}. Note that we do not require the test cases to satisfy the precondition because negative feature vectors from invalid test cases will be filtered out by our learning process. Based on the testing results, we divide $S$ into two disjoint sets. One set includes passed test cases that terminate normally without any memory error or violation of the postcondition, denoted as $S^+$. The other set contains the remaining ones, denoted as $S^-$. Note that we heuristically consider that a test case does not terminate after waiting for a threshold number of time units.
Afterwards, we identify all function calls in $prog$ and add learning points before and after each call. At each learning point $l$, we identify a set of relevant variables, denoted as $V_l$. We apply static program slicing to remove the variables which are visible at $l$ but irrelevant to the postcondition or memory safety. In the example shown in Fig.~\ref{anexample0},
the sets of relevant variables at learning point 1 and 2 are $\{x,n\}$ and $\{x,y\}$ respectively.
For each learning point, we instrument the program to extract a vector of features from each test.

\subsection{Feature Extraction}
Central to our approach is the answer to the question: what features to extract? In this work, we view a program state as a memory graph and systematically extract two groups of features based on the memory graph. One group contains generic features of the memory graph and the other contains features
which are specific to the verification task. Formally, a memory graph $G$ is a tuple $(M, init, E, Ty, L)$ such that

\begin{itemize}
    \item $M$ is a 
      set of heap nodes 
      including a special node \form{\code{null}};
    \item $init \in M$ is a special initial node;
    \item $E$ is a set of labeled and directed edges such that $(s, n, s') \in E$ means that we can access heap node $s'$ via a pointer named $n$ from $s$. An edge starting from $init$ is always labeled with one of the variables in the program.
    \item $Ty$ is a total labeling function which labels each heap node in $M$ by a type;
    \item and $L$ is a labelling function which labels a heap node of primitive type by a value.
\end{itemize}



\begin{wrapfigure}{l}{3.5cm}
\centering
{\ssmall
\begin{tikzpicture}
\tikzset{vertex/.style = {shape=circle,draw,thick,minimum size=2.0em}}
\tikzset{edge1/.style = {->, line width=0.1em}}
\node[vertex] (init) at (0,0)   {$init$};
\node[vertex] (node) at (1.5,0)  {$Node$};
\node[vertex] (null) at (3,1)   {\form{\code{null}}};
\node[vertex] (1) at (3,-1)   {$1$};
\draw[edge1] (init) -- node[midway,below] {$y$} ++ (node);
\draw[edge1] (node) -- node[midway,below] {~~~~~~$next$} ++ (null);
\draw[edge1] (node) -- node[midway,below] {$data$~~~~~~} ++ (1);
\draw[edge1] (init) -- node[midway,above] {$x$} ++ (null);
\end{tikzpicture}}
\caption{A memory graph}
\label{fig:running:example:sampling}
\end{wrapfigure}
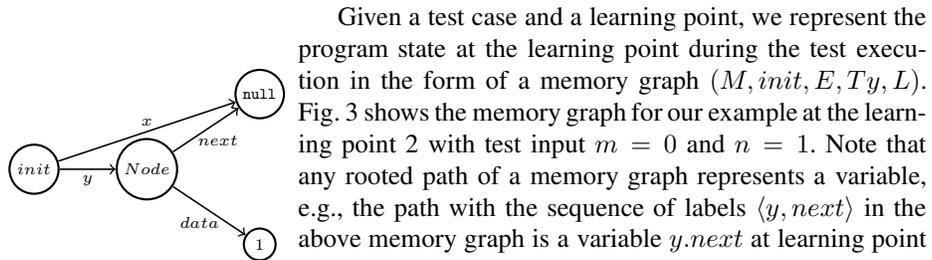
Given a test case and a learning point, we represent the program state at the learning point during the test execution in the form of a memory graph $(M, init, E, Ty, L)$. Fig.~\ref{fig:running:example:sampling} shows the memory graph for our example at the learning point 2 with test input $m=0$ and $n=1$. Note that any  rooted path of a memory graph represents a variable, e.g., the path with the sequence of labels $\langle y, next \rangle$ in the above memory graph is a variable $y.next$ at learning point 2. For complicated programs,
the memory graph might contain many paths and thus many variables from which we can extract features. We thus set a bound on the number of de-referencing to limit the number of variables. For example, if we set the bound to be 2, we focus on variables $\{x, x.data, x.next, y, y.data, y.next\}$ at learning point 2 and similarly variables $\{x, x.data, x.next, n\}$ at learning point 1. With length bounded to 1, we focus only on $\{x, y\}$ at learning point 2 and $\{x, n\}$ at learning point 1.

\begin{table*}[t]
	\ssmall
\begin{center}
\caption{Features}
\label{features1}
\begin{tabular}{| c | c H | c | c H | c | c H |}
\hline
\# & feature & remark & \# & feature & remark & \# & feature & remark \\
\hline
1 & $x=\nil$ & ddd & 10 & $\sepnode{x}{Node}{} {\wedge} \code{is\_sll}(y) {\wedge} \code{sep}(x,y)$ & ddd & 19 & $\code{len\_sll}(x) + \code{len\_sll}(y) > 0$ & \\
2 & $y=\nil$ & ddd & 11 & $\code{is\_sll}(x) {\wedge} \sepnode{y}{Node}{} {\wedge} \code{sep}(x,y)$ & ddd & 20 & $\code{len\_sll}(x) - \code{len\_sll}(y) > 0$ & \\
3 & $\sepnode{x}{Node}{}$ (a.k.a. $x \neq \nil$) & ddd & 12 & $\code{is\_sll}(x) {\wedge} \code{is\_sll}(y) {\wedge} \code{sep}(x,y)$ & ddd & 21 & $- \code{len\_sll}(x) + \code{len\_sll}(y) > 0$ & \\
4 & $\sepnode{y}{Node}{}$ (a.k.a. $y \neq \nil$) & ddd & 13 & $\code{len\_sll}(x) > 0$ & ddd & 22 & $\code{-len\_sll}(x) - \code{len\_sll}(y) > 0$ & \\
5 & $x=y$ & ddd & 14 & $\code{len\_sll}(y) > 0$ &  & 23 & $\code{len\_sll}(x) + \code{len\_sll}(y) = 0$ & \\
6 & $x \neq y$ & ddd & 15 & $\code{len\_sll}(x) < 0$ & & 24 & $\code{len\_sll}(x) - \code{len\_sll}(y) = 0$ & \\
7 & $\code{is\_sll}(x)$ & ddd & 16 & $\code{len\_sll}(y) < 0$ & & 25 & $-\code{len\_sll}(x) + \code{len\_sll}(y) = 0$ & \\
8 & $\code{is\_sll}(y)$ & ddd & 17 & $\code{len\_sll}(x) = 0$ & & 26 & $\code{-len\_sll}(x) - \code{len\_sll}(y) = 0$ & \\
9 & $\sepnode{x}{Node}{} {\wedge} \sepnode{y}{Node}{} {\wedge} \code{sep}(x,y)$ & ddd & 18 & $\code{len\_sll}(y) = 0$ & & & &\\
\hline
\end{tabular}
\end{center}
\end{table*}

We extract two groups of boolean-typed features based on the memory graph. The first group contains generic heap-related features, which include the following.
\begin{itemize}
    \item For each reference type variable $x$, we extract two features which represent if it is \form{\code{null}} or not,
	i.e, whether its corresponding path leads to the special node \form{\code{null}}.
    \item For each pair of reference type variables, we extract two features which represent if the two variables are aliasing or not, i.e., whether their corresponding paths 
	lead to the same non-null node.
    \item For each pair of reference type variables, we extract a feature which represents whether two variables are separated in the memory. Assume that variables $x$ and $y$ lead to nodes $n_x$ and $n_y$, $x$ and $y$ are separated, denoted as \form{\code{sep}(x,y)}, if and only if all reachable nodes except \form{\code{null}} from $n_x$ (including $n_x$) are not reachable from $n_y$ and vice versa.
    \item For each pair of the numeric variables, we extract boolean features in difference logic and the octagon abstract domain~\cite{Mine:2006:OAD:1145489.1145526}, e.g., $\pm x \pm y {>} c$, $\pm x \pm y {=} c$, $\pm x {>} c$ or $x {=} c$ where $c$ is a constant.
	We apply a heuristic to collect constants in conditional expressions in the given program as candidate values for $c$. The value 0 is chosen by default.
\end{itemize}
While general heap-related features are often useful, some programs can only be proven with features which are specific to the verification problem.
Thus, we extract a second group of features based on user-defined predicates used to assert the correctness of the given program, which include the following.
\begin{itemize}
    \item For every permutation of $n$ variables,
	we extract a feature which represents whether the variables satisfy the predicate. For instance, given the user-defined predicate \form{\code{sll}} which has one reference-typed parameter, we extract a feature which represents whether $x$ satisfies the predicate, for each reference variable $x$.
    \item For a pair of two sequences of 
	variables $X$ and $Y$ which satisfy some user-defined predicates, we extract a feature which represents whether the variables are separated in the memory, i.e., all nodes reachable from any variable in $X$ (except \form{\code{null}}) are not reachable from any node in $Y$ and vice versa. This feature is inspired by the separation conjunction operator \form{\sep} in our assertion language. For instance, given $x$ and $y$ which both satisfy \form{\code{is\_sll}}, this feature value is true if and only if all objects in the singly-linked list $x$ and singly-linked list $y$ are disjoint in memory. Note that this feature subsumes the feature \form{\code{sep}(x,y)} explained above.  
    \item For each numerical parameter of the user-defined predicate, we use a variable to represent its value for each sequence of variables which satisfy the predicate. For instance, as \form{\code{sll}} has a numeric parameter, if variable $x$ satisfies \form{\code{sll}}, we use a fresh variable (denoted as \form{\code{len\_sll}} for readability) to represent the value of the numeric parameter. Boolean features of these numeric variables, together with existing numeric variables, are then extracted in the chosen abstract domains.
\end{itemize}
In general, user-defined predicates can be complicated.
Existing heap program verifiers like GRASShopper and HIP maintain a library of commonly used predicates.
We adopt the predicates in their library and define the corresponding functions to extract the above-mentioned features in the form of an extensible library for our approach. Note that this is a one-time effort. 
For instance, Table~\ref{features1} shows the list of 26 features which we extract at learning point 2 for the program shown in Fig.~\ref{anexample0}.

\subsection{Learning for Compositional Verification} \label{learning}
In the following, we present our approach on learning an invariant based on the extracted feature vectors.
Recall that we systematically instrument the program at every learning point, then extract a value for every feature we discussed above. In our implementation, each feature is extracted using a function which returns a boolean value. Afterwards, each test case is executed so that we collect a vector of boolean values (a.k.a.~a feature vector) which represents an abstraction of the memory graph according to the chosen features. If the test case finishes successfully, the feature vector is labeled $positive$; otherwise, it is labeled $negative$. The labeled feature vectors can be organized into a matrix $M$ whose rows are feature vectors and whose columns are the feature values in all test cases. To ensure all feature vectors have the same dimension, if a feature does not apply (e.g., a variable is not accessible in the test case), we set the corresponding feature value to a special default value. For instance, Table~\ref{tbl:before} shows the matrix where the features are sequenced in the same order of Table~\ref{features1}.

\begin{table*}[t]
{\footnotesize
\begin{center}
\caption{Matrix of feature vectors}
\label{tbl:before}
\begin{tabular}{ c c c c c c c c c c c c c c c c c c c c c c c c c c | c}
\multicolumn{27}{c}{\textbf{Vectors obtained from test cases}} \\
1 & 2 & 3 & 4 & 5 & 6 & 7 & 8 & 9 & 10 & 11 & 12 & 13 & 14 & 15 & 16 & 17 & 18 & 19 & 20 & 21 & 22 & 23 & 24 & 25 & 26 & Label \\
\hline
\textbf{1} & \textbf{0} & \textbf{0} & \textbf{1} & \textbf{0} & \textbf{1} & \textbf{1} & \textbf{1} & \textbf{0} & \textbf{0} & \textbf{1} & \textbf{1} & \textbf{0} & \textbf{1} & \textbf{0} & \textbf{0} & \textbf{1} & \textbf{0} & \textbf{1} & \textbf{0} & \textbf{1} & \textbf{0} & \textbf{0} & \textbf{0} & \textbf{0} & \textbf{0} & \textbf{positive} \\
\textbf{1} & \textbf{1} & \textbf{0} & \textbf{0} & \textbf{1} & \textbf{0} & \textbf{1} & \textbf{1} & \textbf{0} & \textbf{0} & \textbf{0} & \textbf{1} & \textbf{0} & \textbf{0} & \textbf{0} & \textbf{0} & \textbf{1} & \textbf{1} & \textbf{0} & \textbf{0} & \textbf{0} & \textbf{0} & \textbf{1} & \textbf{1} & \textbf{1} & \textbf{1} & \textbf{positive} \\
\textbf{0} & \textbf{0} & \textbf{1} & \textbf{1} & \textbf{0} & \textbf{1} & \textbf{1} & \textbf{1} & \textbf{1} & \textbf{1} & \textbf{1} & \textbf{1} & \textbf{1} & \textbf{1} & \textbf{0} & \textbf{0} & \textbf{0} & \textbf{0} & \textbf{1} & \textbf{0} & \textbf{1} & \textbf{0} & \textbf{0} & \textbf{0} & \textbf{0} & \textbf{0} & \textbf{positive} \\
\textbf{0} & \textbf{1} & \textbf{1} & \textbf{0} & \textbf{0} & \textbf{1} & \textbf{1} & \textbf{1} & \textbf{0} & \textbf{1} & \textbf{0} & \textbf{1} & \textbf{1} & \textbf{0} & \textbf{0} & \textbf{0} & \textbf{0} & \textbf{1} & \textbf{1} & \textbf{1} & \textbf{0} & \textbf{0} & \textbf{0} & \textbf{0} & \textbf{0} & \textbf{0} & \textbf{negative} \\
\hline \\
\multicolumn{27}{c}{\textbf{Vectors obtained from memory graph mutation}} \\
0 & 0 & 1 & 1 & 0 & 1 & 1 & 1 & 1 & 1 & 1 & 1 & 1 & 1 & 0 & 0 & 0 & 0 & 1 & 0 & 0 & 0 & 0 & 1 & 1 & 0 & positive \\
0 & 0 & 1 & 1 & 0 & 1 & 0 & 1 & 1 & 1 & 0 & 0 & N & 1 & N & 0 & N & 0 & N & N & N & N & N & N & N & N & negative \\
0 & 1 & 1 & 0 & 0 & 1 & 0 & 1 & 0 & 1 & 0 & 0 & N & 0 & N & 0 & N & 1 & N & N & N & N & N & N & N & N & negative \\
0 & 0 & 1 & 1 & 0 & 1 & 1 & 1 & 1 & 1 & 1 & 1 & 1 & 1 & 0 & 0 & 0 & 0 & 1 & 1 & 0 & 0 & 0 & 0 & 0 & 0 & negative \\
1 & 0 & 0 & 1 & 0 & 1 & 1 & 0 & 0 & 0 & 1 & 0 & 0 & N & 0 & N & 1 & N & N & N & N & N & N & N & N & N & negative \\
0 & 0 & 1 & 1 & 0 & 1 & 1 & 0 & 1 & 0 & 1 & 0 & 1 & N & 0 & N & 0 & N & N & N & N & N & N & N & N & N & negative \\
0 & 0 & 1 & 1 & 1 & 0 & 1 & 1 & 0 & 0 & 0 & 0 & 1 & 1 & 0 & 0 & 0 & 0 & 1 & 0 & 0 & 0 & 0 & 1 & 1 & 0 & negative
\end{tabular}
\end{center}}
\end{table*}

The first step in our learning process is normalising the matrix $M$. 
If there are two rows with the same feature values and same labels, one of them is redundant and removed.
Next, we apply the algorithm in~\cite{DBLP:journals/cacm/Valiant84} to learn a boolean combination of features to separate positive and negative vectors. Informally, the algorithm considers each feature vector as a point in space and every positive point is connected to every negative point by a line. A feature `cuts' a line if the corresponding positive point and negative point have different values for the feature. The goal is to find a list of features that can cut all the lines, i.e., separate all positive and negative points. The features are chosen using a greedy algorithm. At each step, the feature which cuts the most number of uncut lines is selected. After all lines are cut, the selected features partition the space into multiple regions, each of which contains either positive points only or negative points only. Each region can be characterised by a conjunction of the features and the disjunction of all the formulas characterising the positive regions is a boolean formula which separates all the positive and negative feature vectors.

The details are shown in Algorithm \ref{features} and \ref{region}. In Algorithm \ref{features}, the input is a normalised matrix $M$ and the output is the list of features $K$ which can classify all positive and negative rows in $M$. $K$ is initialised as an empty list (line 1). A list $L$ is initialized to contain all pairs of rows $(i,j)$ such that $i$ is the index of a positive row and $j$ is that of a negative row (line 1). During each iteration, the feature $k$ that `cuts' the most number of pairs in $L$ is identified (line 3). 
Note that we do not consider the case $M_{ik} = 0 \wedge M_{jk} = 1$ because it will create the negations of features, which may not be easily transformed into separation logic.
We then remove from $L$ the pairs that are classified correctly by $k$ (line 7) and add the new feature $k$ into $K$ (line 8). The loop stops when $L$ is empty (line 2)
or the best feature at the current iteration cannot classify more pairs (line 4). In the former case, we return the list of features $K$ (line 9). In the latter case, it means the features are not sufficient to distinguish all positive and negative rows. We thus stop and may ask users to provide a new feature (line 5).

\SetAlFnt{\normalsize}

\begin{algorithm}[t]
{
$K=\{\}$; $L=\{(i,j) ~|$ row $i$ is positive and row $j$ is negative$\}$\;
\While{$L$ is not empty}
{
Find $k$ s.t.~$\{(i,j) \in L~|~M_{ik} = 1 \wedge M_{jk} = 0\}$ is the largest\;
\If{the number of pairs $(i,j)$ that $k$ can classify is $0$}{Stop and ask for user input for a new feature\;}
\Else{
Remove $(i,j)$ s.t.~$M_{ik} = 1$ and $M_{jk} = 0$ from $L$\;
Add $k$ to $K$;
}
}
Return $K$\;}
\caption{Choose the list of features $\code{choose}(M)$}
\label{features}
\end{algorithm}

\begin{algorithm}[t]
{
$R=\{\}$; $PP = \{p ~|$ row $p$ is positive$\}$; $NP = \{n ~|$ row $n$ is negative$\}$\;
Mark all $p \in PP$ as uncovered\;
\For{$i = 1$ to $|K|$}
{
Create all combinations $C$ with $i$ elements from the list of features $K$\;
\For{each combination $c \in C$}
{
\If{$\forall n \in NP ~\exists k \in c: M_{nk} = 0$}
{
$CP = \{p ~|~ p \in PP$ and $\forall k \in c: M_{pk} = 1\}$\;
\If{$CP$ contains at least one uncovered index}
{Remove from $R$ the combinations that have the covered indexes are proper subsets of $CP$\;
Add $c$ to $R$; Mark all $p \in CP$ as covered\;
\If{all $p \in PP$ are covered}{Return $R$\;}
}}
}}}
\caption{Combine the features $\code{combine}(M, K)$}
\label{region}
\end{algorithm}

Algorithm \ref{region} then shows how a boolean formula that classifies all positive and negative rows in $M$ is constructed from the chosen features. The input is a normalised matrix $M$ and a list of features $K$ chosen using Algorithm~\ref{features} and the output is a boolean combination of these features. Initially, the list of regions $R$ is empty; $PP$ and $NP$ are the set of indexes of positive and negative rows respectively (line 1). Recall that each row can be seen as a point in space. All points in $PP$ are marked as uncovered at line 2. Favoring simple hypothesis (which is a heuristics often applied in machine learning), we try the combination from 1 feature to $|K|$ (which is the number of features in $K$) features (line 3). At line 4, all the combinations of $i$ features are created. For each combination (line 5), we check if the created region contains no negative points (line 6). If it is the case, we find a list of positive points that are covered by the region (line 7). If this region contains at least one uncovered point (line 8), we add this combination into $R$ and mark the positive points in the region as covered (line 10). Line 9 simplifies the results by removing the chosen regions that only cover a proper subset of positive points in the new region. When all positive points are covered, we return the set of combinations $R$ (line 11 and 12). Each combination is a conjunction of features and the set of combinations is the disjunction of these conjunctions.

For our example, at the learning point 2, after removing redundant rows, we have a matrix with 4 rows and 26 columns, i.e., the bolded rows in Table~\ref{tbl:before}.
Row 1, 2 and 3 are positive, whereas row 4 is negative.
To separate these rows, two columns 1 and 4 are chosen. From this, we can form two regions, in particular, the first one with only column 1, the second one with only column 4.
These two columns represent feature \form{x = \code{null}} (column 1) and \form{y \neq \code{null}} (column 4). As a result, we learn the predicate \form{x = \code{null} \vee y \neq \code{null}}. Note that this predicate is incorrect and it is to be improved later. \\

\noindent It can be shown that Algorithm~\ref{features} and \ref{region} always terminate. The worst-case complexity of Algorithm~\ref{features} is $\mathcal{O}(Row^4 * Col)$ where $Row$ and $Col$ are the number of rows and columns in the input matrix respectively. For Algorithm \ref{region}, the worst-case complexity is $\mathcal{O}(2^{|K|} * (Row * Col + Row^3))$.
While the worst-case complexity is high, these algorithms are often reasonably efficient (as we show in our empirical study). The main reason is that the number of features $K$ (which dominates the overall complexity) is often small (average 1.05 in our experiments).


\subsection{Automatic Memory Graph Mutation} \label{active}
Recall that we only need a correct predicate, which is an invariant at the learning point and sufficient to prove the postcondition.
A fundamental limitation of using classification techniques is
that the learned predicate is likely incorrect if the feature vectors (i.e., test cases) are insufficient. One way to solve this problem is to use a program verifier to check whether the predicate is correct. If it is not correct, the verifier would generate a counterexample and the learning process can continue with a new feature vector obtained from the counterexample. This approach is not ideal for two reasons. One is that verifying heap programs is often costly and thus we would like to avoid it as long as possible. The other is that it is highly nontrivial to construct counterexamples when verifying heap programs~\cite{berdine2012diagnosing}.

Because of that, in this work, to improve the learned predicate, we apply an idea similar in spirit to~\cite{cleve2005locating} to automatically mutate the memory graphs obtained from the test cases
and generate more program states.
For each learned predicate $\constr$, we systematically apply a set of mutation operators based on $\constr$. For each variable $x$ in $\constr$, if it is a reference type, the following mutation operators are applied.
\begin{enumerate}
\item Point $x$ to a freshly constructed object of the right type.
\item Point $x$ to a heap node of the right type in the memory graph (including \form{\code{null}}).
\item Swap $x$ with another reference-type variable.
\end{enumerate}
If $x$ is a primitive type, we follow the idea in \cite{pham2017assertion} and mutate it by setting it to a constant, increasing/decreasing its value with a pre-defined offset, or swapping it with another primitive variable.
The number of mutants we generate depends on the current learned predicate. 

These mutation operators are designed to create states which potentially invalidate the learned predicate. 
For instance, if the current predicate is \form{\code{is\_sll}(x)} $\wedge$ \form{\code{is\_sll}(y)} $\wedge$ \form{\code{sep}(x,y)},
where $x$ and $y$ are two reference variables,
applying the mutation operators allows us to obtain memory graphs which invalidate \form{\code{is\_sll}(x)}, \form{\code{is\_sll}(y)} and/or \form{\code{sep}(x,y)}. The expectation is that such a mutated program state would
lead to violation of the postcondition and thus
be labeled with $negative$. If our expectation is met, the predicate is now more likely to be correct; otherwise, the predicate is incorrect and is refined with the new feature vector.

In the extreme cases when all feature vectors are labeled $positive$ or $negative$, the learned predicates are \form{\code{true}} or \form{\code{false}} respectively. We then apply all mutation operators to all variables at the learning point. In our implementation, the mutation is done automatically by instrumenting statements which mutate the according variables at the learning point. We then run the test suite with the mutated program, collect new feature vectors and new test results. These new feature vectors are added into the matrix to learn new predicates.

The mutation at a learning point in the middle of the program may result in program states which may not be reachable. As a result, the final learned predicate, which is expected to be an invariant, may be weaker than the actual one (if the mutated program state is labeled as $positive$). However, a weaker invariant may still serve our goal of verifying the program. To give an example, in the extreme case, if the postcondition is \form{\code{true}} (and there is no risk of memory error), it is sufficient to learn the invariant \form{\code{true}}. We repeat this process of mutation and learning until the learned invariant converges.

For our example, at the learning point 2, after obtaining the first predicate \form{x = \code{null} \vee y \neq \code{null}},
we apply mutation
and obtain more feature vectors. The new feature vectors are shown in Table~\ref{tbl:before} where \form{\code{N}} is a special value denoting that the feature is not applicable.
Next, applying Algorithm~\ref{features}, the chosen features this time are  \form{x = \code{null}}, \form{\code{is\_sll}(x) \wedge \code{is\_sll}(y) \wedge \code{sep}(x,y)}, \form{\code{len\_sll}(x) < \code{len\_sll}(y)}, and \form{\code{len\_sll}(x) = \code{len\_sll}(y)} (column 1, 12, 21 and 24). From these 4 columns, we form 3 regions: $\{12, 1\}$, $\{12, 21\}$ and $\{12, 24\}$, which are transformed into the invariant \form{\code{inv_2}} we show in Section~\ref{anexample}. Similarly, with the help of state mutation, we improve the learned invariant at $l_1$ from \form{x = \code{null} \vee n > 0} to \form{x = \code{null} \vee (\code{is\_sll}(x) \wedge \code{len\_sll}(x) \leq n)}. 

The process of mutation and learning always terminates. As we only have a finite set of variables and features, the set of feature vectors is finite and thus the process of mutation converges eventually. Furthermore, matrix normalisation guarantees we do not have redundant rows in the matrix and, hence, the matrix is finite and the learning process always terminates.









	













\subsection{Compositional Verification}
\label{veri}
Lastly, we show how we use the learned invariants to verify heap programs in a compositional way. Firstly, we transform each loop in the program into a fresh tail recursive function. Then the loop is replaced with a call to the corresponding function.
Note that in the case of nested loops, 
we create multiple functions in which
the function according to the outer loop will call the function according to the inner loop.
This is a standard strategy adopted from existing program verifiers for heap programs~\cite{Chin:SCP:2012}. We then treat loops in the same way as (recursive) function calls. 

Secondly, we identify the learning points, i.e., before and after each function call statements and learn invariants at these points. Note that we do not learn before/after recursive function calls. This is because program verifiers for heap programs like GRASS-hopper and HIP support inductive reasoning and thus one specification for each recursive function is sufficient. Assume that the invariant learned before function call $C_i$ is $I_i$ and the one learned after $C_i$ is $I_{i+1}$.

Thirdly, for each function call $C_i$, we generate a proof obligation in the form of a Hoare triple $\{I_i\}C_i\{I_{i+1}\}$, to prove that calling function $C_i$ with $I_i$ being satisfied results in a state satisfying $I_{i+1}$. Each proof obligation is submitted to a program verifier. Once the proof obligation is discharged, we replace the function call $C_i$ with its now-established specification, i.e., two statements \form{\code{assert~I_i};} \form{\code{assume~I_{i+1}}}. That is, we instrument the learned invariants into the program such that the invariant learned before/after $C_i$ becomes an assert/assume-statement respectively.

Finally, we use an existing program verifier to verify the transformed program. Note that the program does not contain any function call (other than possibly a recursive call of itself) now.
It is straightforward to see that
the program satisfies the postcondition and is memory-safe with the precondition if all proof obligations are discharged and the transformed program is verified. If any part is not proved and a counterexample is constructed by the verifier,
we use the counterexample to learn new invariants and then try to prove new Hoare triples.



\begin{table*}[tb]
{\scriptsize
\begin{center}
\caption{Results on GRASShopper (Gh)}
\label{tbl:stats2}
\begin{tabular}{| c | c | c | c | c | c | c | c | c | c | c | c |}
\cline{5-12}
 \multicolumn{4}{c|}{} & \multicolumn{2}{c|}{Gh} & \multicolumn{3}{c|}{Gh+\stool} & \multicolumn{3}{c|}{Gh+\stool-Mutation} \\
\hline
Data Structure & Functions & \#Calls & \#Progs & \#V & Time(s) & \#V & Time(s) & L Time(s) & \#V & Time(s) & L Time(s) \\
\hline
Singly-linked & Traverse, Dispose, & 1 & 5 & 5 & 1.50 & 5 & 1.50 & 0 & 5 & 1.50 & 0 \\
    list      & Insert, Remove,    & 2 & 12 & 0 & - & 12 & 4.97 & 202 & 0 & - & 32 \\
			  & Concat            & 3 & 18 & 0 & - & 18 & 10.74 & 610 & 0 & - & 99 \\
\hline
Sorted list & Traverse, Dispose, & 1 & 3 & 3 & 1.40 & 3 & 1.40 & 0 & 3 & 1.40 & 0 \\
            & Insert            & 2 & 6 & 0 & - & 6 & 4.94 & 152 & 4 & 2.71 & 12 \\
			&                   & 3 & 6 & 0 & - & 6 & 6.96 & 368 & 2 & 2.32 & 32 \\
\hline
Binary tree & Traverse, Dispose, & 1 & 3 & 3 & 43.63 & 3 & 43.63 & 0 & 3 & 43.63 & 0 \\
            & Insert            & 2 & 6 & 0 & - & 4 & 90.23 & 134 & 4 & 90.23 & 12 \\
            &                   & 3 & 6 & 0 & - & 2 & 89.26 & 313 & 2 & 89.26 & 30 \\
\hline
\end{tabular}
\end{center}}
\end{table*}

\section{Implementation and Evaluation}  \label{evaluation}
Our approach has been implemented as a prototype, called \stool, with 3070 lines of Java code.
In the following, we evaluate \stool~to answer multiple research questions (RQ). All experiments are conducted on a laptop with one 2.20GHz CPU and 16 GB RAM. To reduce the effect of randomness, we run each experiment 20 times with 10 random test cases each time.\\ 

\noindent \emph{RQ1: Can our approach enhance state-of-the-art verifiers for heap programs?}
We integrate \stool~into two state-of-the-art verifiers for heap programs: GRASShopper and HIP. Although GRASShopper and HIP target the same class of programs, their approaches differ in multiple ways, e.g., they provide a different library of user-defined predicates and they have different verification strategies. They thus allow us to check whether \stool~is general enough to support different program verifiers.
We remark that alternative program verifiers like CPAChecker~\cite{beyer2011cpachecker} and SeaHorn~\cite{DBLP:conf/cav/GurfinkelKKN15} target different classes of programs or program properties and hence are not applicable. The only other tool which is capable of verifying heap programs with heap-related specification is jStar~\cite{distefano2008jstar}, which is, however, no longer maintained.

\begin{table*}[!t]
{\scriptsize
\begin{center}
\caption{Results on HIP}
\label{tbl:stats}
\begin{tabular}{| c | c | c | c | c | c | c | c | c | c | c | c | c |}
\cline{3-13}
 \multicolumn{2}{c|}{} & \multicolumn{3}{c|}{HIP} & \multicolumn{4}{c|}{HIP+\stool} & \multicolumn{4}{c|}{HIP+\stool-Mutation} \\
\hline
Data Structure & Program & Result & \#Succ & Time(s) & Result & \#Succ & Time(s) & L Time(s) & Result & \#Succ & Time(s) & L Time(s) \\
\hline
Singly-linked list & Clean   & Fail & 0 & - & Succ & 20 & 0.37 & 17 & Fail & 0 & - & 3 \\
                   & Clone   & Fail & 0 & - & Succ & 20 & 0.45 & 17 & Fail & 0 & - & 3 \\
                   & Min     & Fail & 0 & - & Fail & 0 & - & 17 & Fail & 0 & - & 2 \\
                   & Reverse & Fail & 0 & - & Fail & 0 & - & 17 & Fail & 0 & - & 3 \\
				   & Sort    & Fail & 0 & - & Fail & 0 & - & 17 & Fail & 0 & - & 3 \\
				   & Insert  & Fail & 0 & - & Succ & 20 & 0.42 & 38 & Fail & 0 & - & 3 \\
				   & Delete  & Fail & 0 & - & Succ & 20 & 0.42 & 37 & Fail & 0 & - & 2 \\
				   & Append  & Fail & 0 & - & Succ & 20 & 0.45 & 90 & Fail & 0 & - & 6 \\
				   & GetLast & Fail & 0 & - & Succ & 20 & 0.42 & 17 & Fail & 0 & - & 3 \\
				   & GetSum  & Fail & 0 & - & Succ & \textbf{15} & 1.02 & 77 & Fail & 0 & - & 6 \\
				   & ToDll   & Fail & 0 & - & Succ & 20 & 0.30 & 17 & Fail & 0 & - & 3 \\
\hline
Doubly-linked list & Clean   & Fail & 0 & - & Succ & 20 & 0.43 & 17 & Succ & 20 & 0.43 & 3 \\
                   & Clone   & Fail & 0 & - & Succ & 20 & 0.67 & 17 & Succ & 20 & 0.67 & 3 \\
                   & Min     & Fail & 0 & - & Fail & 0 & - & 17 & Fail & 0 & - & 3 \\
                   & Reverse & Fail & 0 & - & Fail & 0 & - & 17 & Fail & 0 & - & 3 \\
				   & Sort    & Fail & 0 & - & Fail & 0 & - & 17 & Fail & 0 & - & 3 \\
				   & Insert  & Fail & 0 & - & Succ & 20 & 0.58 & 18 & Succ & 19 & 0.58 & 3 \\
				   & Delete  & Fail & 0 & - & Succ & 20 & 0.65 & 17 & Succ & 20 & 0.65 & 3 \\
				   & Append  & Fail & 0 & - & Succ & 20 & 0.40 & 92 & Fail & 5 & - & 6 \\
\hline
Sorted list        & Clean   & Fail & 0 & - & Succ & 20 & 0.35 & 17 & Succ & 20 & 0.37 & 3 \\
                   & Clone   & Fail & 0 & - & Succ & 20 & 0.37 & 17 & Succ & 18 & 0.35 & 3 \\
				   & Min     & Fail & 0 & - & Succ & 20 & 0.37 & 17 & Succ & 19 & 0.37 & 3 \\
                   & Travel  & Fail & 0 & - & Succ & 20 & 0.54 & 17 & Succ & 18 & 0.54 & 2 \\
				   & Insert  & Fail & 0 & - & Fail & 0 & - & 16 & Fail & 0 & - & 3 \\
				   & Delete  & Fail & 0 & - & Fail & 0 & - & 18 & Fail & 0 & - & 3 \\
\hline
Cycle list         & Clean   & Fail & 0 & - & Fail & 0 & - & 17 & Fail & 0 & - & 3 \\
                   & Min     & Fail & 0 & - & Fail & 0 & - & 17 & Fail & 0 & - & 3 \\
				   & Travel  & Fail & 0 & - & Succ & 20 & 0.30 & 17 & Fail & 0 & - & 3 \\
                   & ToSll   & Fail & 0 & - & Fail & 0 & - & 17 & Fail & 0 & - & 3 \\
\hline
Binary tree & InOrder   & Fail & 0 & - & Succ & 20 & 0.43 & 16 & Succ & 20 & 0.43 & 2 \\
                   & PreOrder  & Fail & 0 & - & Succ & 20 & 0.46 & 17 & Succ & 20 & 0.46 & 3 \\
                   & PostOrder & Fail & 0 & - & Succ & 20 & 0.45 & 17 & Succ & 20 & 0.45 & 3 \\
                   & Min       & Fail & 0 & - & Succ & 20 & 0.51 & 17 & Succ & 20 & 0.51 & 3 \\
				   & Max       & Fail & 0 & - & Succ & 20 & 0.51 & 17 & Succ & 20 & 0.51 & 3 \\
				   & Prec      & Fail & 0 & - & Succ & 20 & 0.57 & 17 & Succ & 20 & 0.57 & 3 \\
				   & Succ      & Fail & 0 & - & Succ & 20 & 0.57 & 17 & Succ & 20 & 0.57 & 3 \\
				   & Insert    & Fail & 0 & - & Succ & 20 & 0.67 & 17 & Succ & 20 & 0.67 & 3 \\
				   & Delete    & Fail & 0 & - & Fail & 0 & - & 22 & Fail & 0 & - & 3 \\
\hline
AVL tree           & Insert    & Fail & 0 & - & Fail & 0 & - & 17 & Fail & 0 & - & 3 \\
                   & Delete    & Fail & 0 & - & Fail & 0 & - & 24 & Fail & 0 & - & 3 \\
\hline
Red-black tree     & Insert    & Fail & 0 & - & Fail & 0 & - & 22 & Fail & 0 & - & 3 \\
                   & Delete    & Fail & 0 & - & Fail & 0 & - & 38 & Fail & 0 & - & 3 \\
\hline
MCF                & Travel    & Fail & 0 & - & Fail & 0 & - & 17 & Fail & 0 & - & 3 \\
\hline
Rose tree          & Travel    & Fail & 0 & - & Fail & 0 & - & 17 & Fail & 0 & - & 3 \\
\hline
Tll                & SetRight  & Fail & 0 & - & Succ & 20 & 2.40 & 16 & Succ & 19 & 2.40 & 2 \\
\hline
\end{tabular}
\end{center}}
\end{table*}

We conduct two sets of experiments based on these two verifiers. Our first experiment is with GRASShopper. Although GRASShopper supports inductive predicates for describing data structures, unlike HIP, it does not support reasoning about separation logic directly. The inductive predicates in GRASShopper are defined based on
first-order logic with some built-in predefined predicates. Due to GRASShopper's limitation, we conduct an experiment based on a set of benchmark programs in its distribution. All programs and experimental results are available at~\cite{data} and the tool is available at~\cite{benchmark}.

The GRASShopper distribution contains many functions for different types of data-structures. We focus on those non-trivial recursive functions with precondition and postcondition. To check how GRASShopper performs with and without \stool, we generate a set of composite programs which randomly invoke one or more of these functions. The function call sequence is formed such that the postcondition of a previous function is identical
(via syntactical checking) to the precondition of the subsequent function. The precondition of the composite program is composed from preconditions of invoked functions and the postcondition of the last function in the call sequence is the postcondition of the composite program.   
In total, we generate 65 composite programs containing 1, 2 and 3 function calls.

Table~\ref{tbl:stats2} shows the results, where the first four columns show the type of data structure, the involved functions, the number of function calls and the number of programs in the category. The next column shows the result of GRASShopper without the help of \stool, i.e., the program is verified using GRASShopper without the specification of each invoked function in the program. We measure the number of verified programs (column \#V) and the time taken. The next column shows the results of GRASShopper enhanced with \stool. No additional user-defined predicates besides those provided in GRASShopper are used in our experiments.
Note that we extract features automatically based on the user-defined predicates in GRASShopper in the experiment.

Without \stool, GRASShopper only verifies 11 (out of 65) programs with 1 function call. 
For the remaining 54 programs which have 2 or 3 function calls, GRASS-hopper fails to verify any of them. This is expected as GRASShopper is unable to derive the necessary function specification automatically.
Enhanced with \stool, GRASShopper verifies 59 (out of 65) programs. For all these programs, we learn the correct invariants in every one of the 20 runs.

The second experiment is with HIP.
We generate 45 programs based on common operations for 10 different data structures.
Each program consists of multiple function calls. Each program starts with a call of a constructor which creates an object of the target data structure (e.g., a singly-linked list), or a function which reads the data structure (e.g., checking whether the root node is $\code{null}$, or traveling through the data structure). Lastly, a function supported by HIP for this data structure is called which may modify the data structure. The postcondition of the program is the postcondition of the last function. The precondition is manually written and checked to guarantee that the program terminates and satisfies the postcondition without any memory error.

Table~\ref{tbl:stats} shows the results, where column $Program$ shows the last
function called in the program. Column \emph{HIP+\stool} shows the results using HIP enhanced with \stool. Note that
we may not be able to learn the same invariants every time due to randomness in generating the initial set of test cases. Thus, we add a column $\#Succ$ to show how many times, out of 20, we are able to learn the invariant and verify the program. No additional user-defined predicates besides those defined in HIP are used in our experiments.
Column \emph{HIP} shows that without \stool, none of these programs is verified. With \stool, HIP successfully verifies 27 programs. In all but 1 case (highlighted with bold) we are able to learn the same invariant consistently.\\
\noindent \emph{RQ2: Which features are useful in verifying heap programs?} We learn invariants based on two groups of features, i.e., general heap-related features and those specific to user-defined predicates. The question is whether these two groups of features are useful and whether there are other features which we could learn based on.

In total, \stool~learned 104 invariants 
(74 
with GRASShopper and 30 
with HIP)
to help solving the verification tasks.
Among them, 93 invariants 
(66 
with GRASShopper and 27
with HIP)
contain only features extracted based on the user-defined predicates (e.g., $ds(x)$ or $ds(x) \sep ds(y)$ with $ds$ being a user-defined predicate). The remaining 11 invariants are additionally constituted with generic features (e.g., $x = \nil$ or $x \neq \nil$). None of the invariants is constituted with general heap-related features only. The results show that the user-defined predicates are important and invariants specific to a verification problem are needed for proving the program. 
Generic heap-related features are also necessary sometimes (in 11\% of the cases).

A total of 24 programs (6 with GRASShopper and 18 with HIP) are not verified.
There are two main reasons why they cannot be proved
even with the help of \stool.
Firstly, some programs can only be verified with complex function specifications which require features that are not supported in \stool.
For example, to prove the remaining 6 programs in the experiment with GRASShopper, we need a feature
characterizing the paths in the tree, which cannot be derived from user-defined predicates.
This is similarly the case for experiments with HIP. One remedy is to extend our implementation with additional features through automatic lemma learning~\cite{Le:TACAS:2018}.
Secondly, there are programs that have
a hierarchy of function calls, e.g., function calls within recursive functions.
Some of the function calls occur under strict condition which is never satisfied by the test cases and thus we are unable to learn the specification of those function calls. This is a fundamental limitation of dynamic analysis approaches, which could be overcome with a comprehensive test suite from a systematic test case generation approach~\cite{Pham17fm,DBLP:journals/corr/Pham17,Pham:2018:THP:3183440.3194964}.\\

\noindent \emph{RQ3: Is memory graph mutation helpful?}
We compare the performance of the enhanced GRASShopper and HIP with and without memory graph mutation. The results are shown in the last columns of Table~\ref{tbl:stats2} and~\ref{tbl:stats}. It can be observed that without memory graph mutation, the number of verified programs by GRASShopper is reduced from 59 to 23, and the number of verified programs by HIP is reduced from 27 to 17. It thus clearly shows that memory graph mutation helps to improve the correctness of the learned invariants. Furthermore, we observe that without memory graph mutation, it is more likely that different invariants are learned in different runs of the same experiments (refer to column $\#Succ$). This is expected as without memory graph mutation, we cannot discard invariants which are the result of limited test cases.\\

\noindent \emph{RQ4: What is the overhead of invariant generation?}
We measure the time taken to learn the invariants.
Columns $L~Time$ in Table~\ref{tbl:stats2} and~\ref{tbl:stats} show the results.
In general, the learning time depends on the number of learning points, the complexity of the program and the initial test suite. Overall, the time required for learning is reasonable, ranging from seconds to minutes.
In the most time consuming case, we spent 92 seconds to learn two invariants for program ``doubly-linked list append''.
For most of the cases, the learning time is about 20 seconds.\\

\noindent \emph{RQ5: Does our invariant generation approach complement existing ones?} The most noticeable invariant generation tool for heap program is Infer~\cite{infer}.
However, Infer is not designed to support verification task. Instead, it generates generic specifications to capture the footprints of the pointers used in the functions based on bi-abduction. We apply Infer to generate specifications (e.g., pre/postconditions) for every function experimented above and notice that they are too weak for program verification. \\


\noindent \emph{Threats to Validity} Firstly, the set of programs used in our experiments are limited compared to real-world data-structure libraries. This is because state-of-the-art verifiers for heap programs are still limited to relatively simple heap programs due to the great difficulty in verifying heap programs. As our experiments show, \stool~successfully enhances the capability of state-of-the-art heap program verifiers so that programs with multiple functions can be automatically verified. Secondly,
\stool~only works when we have the right features in the learning process.
  We expect that applying lemma synthesis could help us obtain more features and overcome this limitation.

\section{Related Work}
The closest to our work is approach for invariant inference using dynamic 
analysis with separation
logic abstraction
\cite{DBLP:journals/corr/abs-1903-09713}.
Similar to our work, it generates invariant based on user-defined predicates (i.e., features in our work).
In contrast to ours, it made use of positive features only
and did not support mutation.
Close to our work are proposals for automatic program verification
using
black-box techniques adopted from the machine learning community.
In particular,
the method presented in \cite{Zhu:PLDI:2016} is based on
user-supplied templates. It is designed to learn specification
for heap programs which ensures no memory errors.
The approach in~\cite{DBLP:journals/corr/LiTBZ15} proposes to learn features from graph-structured inputs
based on neural networks.
The authors showed an application on verifying memory safety using the learning results.
In contrast to~\cite{DBLP:journals/corr/LiTBZ15},
our goal is to learn invariants
to compositionally verify the program against a given specification as well as ensure no memory errors.
In~\cite{Kulkarni:OOPSLA:2016}, the authors presented a method
to learn shared module codes and reuse them during an analysis.
The work in~\cite{Garg:CAV:2013} builds polynomial time active learning algorithms for automaton model of array and list structures.
Our proposal also relies on a learning algorithm and actively improves the learned invariants. 
In~\cite{Muhlberg:SEFM:2015}, the authors proposed a learning method targeted 
lists only.
This method learns the sequence of actions (remove or insert)
from a program and infers the data structures
manipulated by the program. However, it is hard to extend the method to support arbitrary heap-based programs.
Similarly to ours, \cite{Brockschmidt:SAS:2017} guesses
invariants from concrete program states
and checks them by a theorem prover.
However, their work only focuses on list-based programs.
The ICE method proposed in~\cite{Garg:CAV:2014,Garg:POPL:2016} supports inductive properties of loop invariant learning. Besides using the positive and negative points,
ICE proposes additional implication points
to encode the inductive checking for learning invariant.
It is our future work to integrate the idea of ICE learning with our graph-based learning. The work in~\cite{Padhi:PLDI:2016} presents an approach for precondition inference. The main contribution is feature learning for functional programs.
It is interesting to apply the feature learning techniques in our future work.

Our work is also related to automatic and static analyzers for the shape analysis problems, e.g., TVLA~\cite{Sagiv:POPL:1999} and separation logic~\cite{Calcagno:JACM:2011,Chin:SCP:2012,Dudka:CAV:2011,Holik:CAV:2013,Loc:CAV:2014}, and for the verification problem
 of programs that requires both heap and data reasoning, e.g., 
 PDR \cite{Itzhaky:CAV:2014}, interpolation \cite{Aws:ESOP:2015} and
template-based invariant generation \cite{Malik:FMCAD:2018}.
To infer shape-based specification, while tools
 \cite{Calcagno:JACM:2011,Dudka:CAV:2011,Loc:CAV:2014} are based on
the bi-abduction technique,
we use machine learning to obtain a generalized invariant
from a set of concrete executions.
In our implementation, we use GRASSHopper and HIP as external verification engines. As our approach is independent from the program verifiers, we plan to build a general framework so that different verifiers can be used.
Lastly, this work is related to previous works on invariant generation, e.g., Daikon \cite{ernst2007daikon}, or Houdini \cite{flanagan2001houdini}.
However, those works do not focus on learning invariants related to data structures like this one.

 \label{related}
\section{Conclusion} \label{conclusion}
We have presented a novel learning approach
to the automated and compositional verification of
heap-manipulating programs.
The essence of our approach is an algorithm
to infer invariants based on a set of memory graphs representing the program states obtained
from concrete executing traces.
We further enhance the precision of learned invariant
with memory graph mutation.
We have implemented a prototype tool
and evaluated it over a set of programs which
manipulate complex data structures.
The experimental results show that
our tool enhances the capability of existing program verifiers
to 
verify nontrivial heap-based programs.
In the future, we 
might apply our tool to more verifiers and more test subjects
as well as compare our tool with other tools, e.g., Predator \cite{Dudka:CAV:2011}, Forester \cite{10.1007/978-3-662-54580-5_24,Holik:CAV:2013},  S2 \cite{Loc:CAV:2014}, and SLING \cite{DBLP:journals/corr/abs-1903-09713}.


\bibliographystyle{splncs04}
\bibliography{arf}
\end{document}